\documentclass[a4paper,11pt]{article}

\pdfoutput=1

\usepackage{amsmath,amssymb,mathtools}
\usepackage{color}
\usepackage{graphicx}
\usepackage{subfigure}
\usepackage{cite}
\usepackage{hyperref}
\usepackage{multirow,makecell}
\usepackage{textcomp}
\usepackage{wasysym}

\usepackage{amsthm}

\usepackage{ulem} %% package for delete line \sout \xout

\theoremstyle{definition}

\usepackage[text={17cm,24.5cm},centering]{geometry} %% thanks to Chao Wu %% full A4 size is 21cm*29.7cm
\numberwithin{equation}{section} %%equastion number with section number

\newcommand{\mlr}[2]{\multirow{#1}*{#2}} %%\multirow
\newcommand{\mlc}[3]{\multicolumn{#1}{#2}{#3}} %%\multicolumn

\def \be {\begin{equation}}
\def \ee {\end{equation}}
\def \ba {\begin{array}}
\def \ea {\end{array}}
\def \bea{\begin{eqnarray}}
\def \eea{\end{eqnarray}}
\def \nn {\nonumber}

\def \b {\beta}
\def \g {\gamma}

\def \d {\delta}
\def \D {\Delta}
\def \e {\epsilon}

\def \m {\mu}

\def \l {\lambda}

\def \s {\sigma}

\def \r {\rho}
\def \o {\omega}
\def \O {\Omega}

\def \mB {\mathcal B}

\def \mX {\mathcal X}

\def \cJ {{\mathcal J}}

\def \rL {{\mathrm L}}
\def \rM {{\mathrm M}}

\def \rS {{\mathrm S}}

\def \p {\partial}

\def \f {\frac}

\def \lt {\left}
\def \rt {\right}

\def \sr {\sqrt}
\def \td {\tilde}

\def \inf {\infty}

\def \lag {\langle}
\def \rag {\rangle}

\def \ep {\mathrm{e}}
\def \ii {\mathrm{i}}

\def \tr {\textrm{tr}}

\def \and {{~\textrm{and}~}}

\def \CE {{\rm{CE}}}
\def \ME {{\rm{ME}}}
\def \PE {{\rm{PE}}}

\def \rL {{\textrm{L}}}

\begin{document}

%\title{\textbf{Distinguishably of canonical ensemble, microcanonical ensemble, and primary excited states in 2D CFT}}
\title{\textbf{R\'enyi entropy at large energy density in 2D CFT}}
\author{
Wu-zhong Guo$^{1}$\footnote{wzguo@cts.nthu.edu.tw}~,
Feng-Li Lin$^{2}$\footnote{%Corresponding author.
fengli.lin@gmail.com}~
and
Jiaju Zhang$^{3}$\footnote{%Corresponding author.
jzhang@sissa.it}
}
\date{}

\maketitle

\vspace{-10mm}
\begin{center}
{\it
$^{1}$Physics Division, National Center for Theoretical Sciences, National Tsing Hua University,\\
No.\ 101, Sec.\ 2, Kuang Fu Road, Hsinchu 30013, Taiwan\\\vspace{1mm}
$^{2}$Department of Physics, National Taiwan Normal University,\\
No.\ 88, Sec.\ 4, Ting-Chou Road, Taipei 11677, Taiwan\\\vspace{1mm}
$^{3}$SISSA and INFN, Via Bonomea 265, 34136 Trieste, Italy
}
\vspace{10mm}
\end{center}

\begin{abstract}
  We investigate the R\'enyi entropy and entanglement entropy of an interval with an arbitrary length in the canonical ensemble, microcanonical ensemble and primary excited states at large energy density in the thermodynamic limit of a two-dimensional large central charge $c$ conformal field theory. As a generalization of the recent work [Phys. Rev. Lett. 122 (2019) 041602], the main purpose of the paper is to see whether one can distinguish these various large energy density states by the R\'enyi entropies of an interval at different size scales, namely, short, medium and long. Collecting earlier results and performing new calculations in order to compare with and fill gaps in the literature, we give a more complete and detailed analysis of the problem. Especially, we find some corrections to the recent results for the holographic R\'enyi entropy of a medium size interval, which enlarge the validity region of the results. Based on the R\'enyi entropies of the three interval scales, we find that R\'enyi entropy cannot distinguish the canonical and microcanonical ensemble states for a short interval, but can do the job for both medium and long intervals. At the leading order of large $c$ the entanglement entropy cannot distinguish the canonical and microcanonical ensemble states for all interval lengths, but the difference of entanglement entropy for a long interval between the two states would appear with $1/c$ corrections. We also discuss R\'enyi entropy and entanglement entropy differences between the thermal states and primary excited state. Overall, our work provides an up-to-date picture of distinguishing different thermal or primary states at various length scales of the subsystem.
\end{abstract}

\baselineskip 18pt
\thispagestyle{empty}

\newpage

\tableofcontents

%\begin{center}
%\textbf{\LARGE On single interval entanglement entropy and relative entropy from OPE of twist operators}
%\end{center}

\section{Introduction}

Eigenstate thermalization hypothesis (ETH) \cite{Deutsch:1991,Srednicki:1994,Srednicki:1995pt,rigol2008thermalization,DAlessio:2016rwt} states that in a chaotic system local operators cannot distinguish a highly excited energy eigenstate from a proper thermal state.
Then, the natural and complementary questions are whether some nonlocal measures can distinguish the excited and thermal states, and how nonlocal they should be.
A natural set of nonlocal operators are the entanglement entropy $S_A $ and R\'enyi entropy $S_A^{(n)}$ of a subsystem $A$ of volume $V_A$ in a system of volume $V$ and state with density matrix $\rho$.
The reduced density matrix of $A$ is obtained by tracing out the degrees of freedom of its complement $\bar A$, i.e.  $\rho_A=\tr_{\bar A}\r$.
The R\'enyi entropy is defined as
\be
S_A^{(n)} = - \f{1}{n-1} \log\tr_A \r_A^n,
\ee
and in $n \to 1$ limit it becomes the entanglement entropy
\be
S_A = - \tr_A ( \r_A \log \r_A ).
\ee
Based on these and other nonlocal quantities the subsystem ETH was proposed \cite{Lashkari:2016vgj,Dymarsky:2016ntg,Lashkari:2017hwq}.  Moreover, the distinguishability of a thermal state from its microstates is related to the black hole information loss paradox \cite{Hawking:1974sw,Hawking:1976ra} through gauge/gravity duality \cite{Maldacena:1997re,Gubser:1998bc,Witten:1998qj}.

The above scheme of distinguishability can be extended to the states with finite energy density,  i.e. states of energy $E$ with $E/V$  fixed and finite in the thermodynamic limit  $V \to \inf$. For these states the R\'enyi entropy is expected to follow the volume law \cite{Lu:2017tbo}.
A related question is whether entanglement entropy or R\'enyi entropy can distinguish canonical and microcanonical ensemble states even in the thermodynamic limit.
As local operators cannot distinguish the canonical and microcanonical ensemble states \cite{Dymarsky:2016ntg,Lashkari:2017hwq,Faulkner:2017hll,Guo:2018djz}, neither can the short interval entanglement entropy or R\'enyi entropy.
It was proposed in \cite{Dymarsky:2016ntg} that the R\'enyi entropy of an interval with a length that is comparable to the length of its complement can distinguish the canonical and microcanonical ensemble states, while the entanglement entropy cannot.
Recently, in this context it was shown by Dong in \cite{Dong:2018lsk} (which is motivated by \cite{Lu:2017tbo} and \cite{Garrison:2015lva,Dymarsky:2016ntg}) that the holographic R\'enyi entropy of a medium size subsystem, i.e. with $V_A/V$ fixed and finite, can distinguish the canonical and microcanonical ensemble states of large energy density, i.e. $E/V \propto c$ with $c$ being the large central charge. Holographically, one can evaluate the entanglement entropy by Ryu-Takayanagi formula \cite{Ryu:2006bv,Hubeny:2007xt,Casini:2011kv,Lewkowycz:2013nqa,Dong:2016hjy}, and the R\'enyi entropy by relating it to the refined R\'enyi entropy \cite{Dong:2016fnf}
\be
\td S_A^{(n)} = n^2 \p_n \Big( \f{n-1}{n} S_A^{(n)} \Big),
\ee
which can be evaluated by the area of a bulk codimension-two cosmic brane.

 Motivated by \cite{Dong:2018lsk}, in this paper we investigate R\'enyi entropy in two-dimensional (2D) CFTs with a large central charge $c$ in various states of large energy density in the thermodynamic limit.
The length of the circle where the CFT lives is $L$, and the length of the subsystem $A$ is $\ell$.
For comparison, we consider three types of large density states:
(i) the canonical ensemble state at inverse temperature $\beta$;
(ii) the microcanonical ensemble state of energy $E=\f{\pi c L}{6\l^2}$ with $\l$ being a constant;
and (iii) a primary excited  state of  conformal weights $h=\bar h =\f{c}{24} ( \f{L^2}{\m^2} + 1 )$, or, equivalently, energy $E=\f{\pi c L}{6\m^2}$ and spin $s=0$, with constant $\m$.
Furthermore, for each kind of states, we will consider three different  scales of $\ell$.
For the canonical ensemble state, we have
(S) short interval with $0<\ell \lesssim \beta$;
(M) medium interval with $\beta \lesssim \ell \lesssim L-\beta$;
and (L) long interval with $L-\beta \lesssim \ell \lesssim L$.
In the above we have used ``$\lesssim$'' to indicate that we cannot find the sharp regime boundaries.
It is similar for the microcanonical ensemble state.
For the primary excited state, we have
(S) short interval with $0<\ell \lesssim \m$;
(M) medium interval with $\m \lesssim \ell < L/2$;
and (L) long interval with $L/2 < \ell \lesssim L$.

The medium interval regime for canonical and microcanonical ensemble states was recently investigated in \cite{Dong:2018lsk} for holographic CFTs in general dimensions, and the 2D results for holographic R\'enyi entropy and refined R\'enyi entropy in this regime are
\bea \label{ScenMSmeMnD}
&& S_{\CE,\rM}^{(n)}(\ell) = \f{\pi c (n+1) \ell}{6n\b}, ~~
   \td S_{\CE,\rM}^{(n)}(\ell) = \f{\pi c \ell}{3 n \b}, \nn\\
&& S_{\ME,\rM}^{(n)}(\ell) = \f{\pi c nL}{3(n-1)\l} \Big( 1-  \sr{1-\f{\ell}{L}+\f{\ell}{n^2L}} \Big), \nn\\
&& \td S_{\ME,\rM}^{(n)}(\ell) = \f{\pi c \ell}{3n\l\sr{1-\f{\ell}{L}+\f{\ell}{n^2L}}}.
\eea
In the above equation we have used ``CE'' to denote the canonical ensemble state and ``ME'' to denote the microcanonical ensemble state, and later in this paper we will also use ``PE'' to denote the primary excited state.
It was argued in \cite{Lu:2017tbo,Dong:2018lsk} that the above results for microcanonical ensemble state also apply to the primary excited state as long as $\ell < L/2$. Note that, the primary excited state is a pure state so that $S_\PE^{(n)}(\ell)=S_\PE^{(n)}(L-\ell)$ relating the short and long interval ones, and the short interval expansion has been obtained in \cite{Lashkari:2016vgj,Lin:2016dxa,He:2017vyf,He:2017txy}.
 %Besides the above, in this paper we need to do the calculations for other cases.
 R\'enyi entropy in canonical ensemble state for all three scales of  $\ell$ have been obtained in  \cite{Calabrese:2004eu,Chen:2017ahf} by field theory method.
However, R\'enyi entropy in microcanonical ensemble haven't been explored  before. This is one of the concrete results of our paper.
 Combining all the results, we obtain the piecewise R\'enyi entropies in canonical ensemble, microcanonical ensemble and primary excited states for arbitrary subsystem size.

Using these piecewise R\'enyi entropies, we find that the short interval R\'enyi entropy cannot distinguish a canonical ensemble state from a microcanonical ensemble one as expected, but the medium and long interval ones can.
In contrast, {at the leading order of $c$} the entanglement entropy of any length $\ell$ cannot distinguish the canonical and microcanonical ensemble states.
With the $1/c$ corrections, however, the entanglement entropy of long interval regime can distinguish these two ensembles.
Our findings are consistent with and generalize the holographic results in \cite{Dong:2018lsk}.
The results are summarized in Table~\ref{tab}.

In the calculations we will use the powerful method of twist operators \cite{Calabrese:2004eu,Cardy:2007mb} and their operator product expansion (OPE) \cite{Headrick:2010zt,Calabrese:2010he,Rajabpour:2011pt,Chen:2013kpa} to calculate the R\'enyi entropies of 2D CFTs for short interval \cite{Chen:2016lbu,Lin:2016dxa,He:2017vyf,He:2017txy} and long interval \cite{Chen:2014ehg,Chen:2015kua}.
The result for canonical ensemble state in \cite{Chen:2017ahf} will also be useful to us.

 The remaining part of the paper is arranged as follows.
In sections~\ref{secCE}, \ref{secME}, \ref{secPE}, we investigate the R\'enyi entropies in, respectively, the canonical ensemble state, the microcanonical ensemble state and primary excited state.
We investigate the distinguishabilities of the various states by the R\'enyi entropy and entanglement entropy in section~\ref{secdist}.
We conclude with discussion in section~\ref{secdisc}.
In appendix~\ref{appMI}, we review the R\'enyi mutual information of two intervals on a plane in the ground state.
In appendix~\ref{appdetails}, we collect some calculation details in section~\ref{secME}.

\section{Canonical ensemble state} \label{secCE}

 Most of the results in this section are not new, and we collect the results in literature for completeness and comparison.

We first consider the canonical ensemble state in 2D CFT at inverse temperature $\b$. R\'enyi entropy of a length $\ell$ interval in the thermodynamic limit is well-known \cite{Calabrese:2004eu}
\be \label{SceS}
S_{\CE,\rS}^{(n)}(\ell) = \f{c(n+1)}{6n} \log \Big( \f{\b}{\pi\e} \sinh \f{\pi\ell}{\b} \Big),
\ee
with $\e$ being the UV cutoff.
Note that this formula holds as long as $ \ell \lesssim L - \b$, and was obtained by using the method of twist operators \cite{Calabrese:2004eu,Cardy:2007mb}. The long interval R\'enyi entropy has also been investigated in \cite{Chen:2014ehg,Chen:2015kua} and \cite{Chen:2017ahf}, and we just adopt the result in \cite{Chen:2017ahf} that was obtained using conformal transformations. In the thermodynamic limit, the result is
\be \label{SceL}
S_{\CE,\rL}^{(n)}(\ell) = \f{c(n+1)}{6n} \log \Big( \f{\b}{\pi\e} \sinh \f{\pi(L-\ell)}{\b} \Big)
                        + \f{\pi c(n+1)L}{6n\b}
                        - I_n\big(1-\ep^{-\f{2\pi(L-\ell)}{\b}}\big).
\ee
This formula holds as long as $\ell \gtrsim \b$. In the above, we have introduced $I_n(x)$ with $0<x<1$, which is the R\'enyi mutual information of two disjoint intervals on a complex plane with cross ratio $x$,  see a brief review in Appendix~\ref{appMI}.  Note that $I_n(x)$ satisfies the property \cite{Headrick:2010zt}
\be
I_n(x) = \f{c(n+1)}{6n}\log\f{x}{1-x} + I_n(1-x).
\ee
For $x \ll 1$ it has been calculated up to order $x^{8}$ \cite{Hartman:2013mia,Faulkner:2013yia,Barrella:2013wja,Chen:2013kpa,Chen:2013dxa}. In this paper we use the small $x$ expansion up to $x^{8}$ as the approximation of $I_n(x)$ for $0<x<1/2$.
When $\ell$ approaches $L-\e$, the long interval R\'enyi entropy approaches the R\'enyi entropy of the entire system
\be \label{Scetot}
S_{\CE}^{(n)}(L) = \f{\pi c(n+1)L}{6n\b}.
\ee

Both (\ref{SceS}) and (\ref{SceL}) work for the medium interval, i.e,  $\b \lesssim \ell \lesssim L-\b$. From either of them we get the medium interval R\'enyi entropy in the thermodynamic limit
\be \label{SceM}
S_{\CE,\rM}^{(n)}(\ell) = \f{c(n+1)}{6n} \log \f{\b}{2\pi\e} + \f{\pi c (n+1) \ell}{6n\b}.
\ee
The medium interval refined R\'enyi entropy is
\be \label{tdSceM}
\td S_{\CE,\rM}^{(n)}(\ell) = \f{c}{3n} \log \f{\b}{2\pi\e} + \f{\pi c \ell}{3n\b}.
\ee
Comparing (\ref{tdSceM}) with the holographic result in \cite{Dong:2018lsk}, i.e. $\td S_{\CE,\rM}^{(n)}(\ell)$ in (\ref{ScenMSmeMnD}), we find an extra term, i.e. the first term on the RHS of (\ref{tdSceM}).
Holographically, the refined R\'enyi entropy is given by the area of a codimension-2 cosmic brane homologous to the interval $A$ in a backreacted bulk geometry, which is denoted by $\mB_n(\b,A)$ in \cite{Dong:2018lsk}.
The second term on the RHS of (\ref{tdSceM}), is the contribution from the part of the cosmic brane parallel to the black hole horizon, and the first term is the part extending along the radial direction of the bulk geometry.
For the second term to dominate the RHS of (\ref{tdSceM}), we need
\be \label{req1}
\f{\ell}{\b} \gg \log\f{\b}{\e}.
\ee
This is consistent with the validity regime of the result in \cite{Dong:2018lsk}
\be \label{req2}
\f{L}{\b} \gg \log\f{\b}{\e}.
\ee
Note that for a medium interval case, by construction $L$ are at the same scale as $\ell$, i.e.,  $\ell\sim L$ so that in the thermodynamic limit we have $\ell/\b \gg 1$, $L/\b \gg 1$.
The requirements (\ref{req1}) and (\ref{req2}) are equivalent for a medium interval in the thermodynamic limit.
For the validity of (\ref{SceM}) and (\ref{tdSceM}), we do not require (\ref{req1}) or (\ref{req2}).
The results (\ref{SceM}) and (\ref{tdSceM}) are generalizations of the results in \cite{Dong:2018lsk} with a larger validity regime.

Due to the aforementioned regions of validity for the short and long interval formulas (\ref{SceS}) and (\ref{SceL}), we can infer that the medium interval formula (\ref{SceL}) should also exist a validity region, say $\ell_1^\CE \lesssim \ell \lesssim \ell_2^\CE$ for some critical lengths $\ell_{1,2}^\CE$.
We cannot determine the precise values of the critical lengths and just adopt the approximate values
\be
\ell_1^\CE \approx \b, ~~ \ell_2^\CE \approx L - \f{\b}{2\pi}\log 2.
\ee
Note that the above $\ell_2^\CE$ is just the critical point for the minimal surface of the holographic entanglement entropy \cite{Azeyanagi:2007bj,Blanco:2013joa,Hubeny:2013gta}.

\begin{figure}[htpb]
  \centering
  \includegraphics[width=0.9\textwidth]{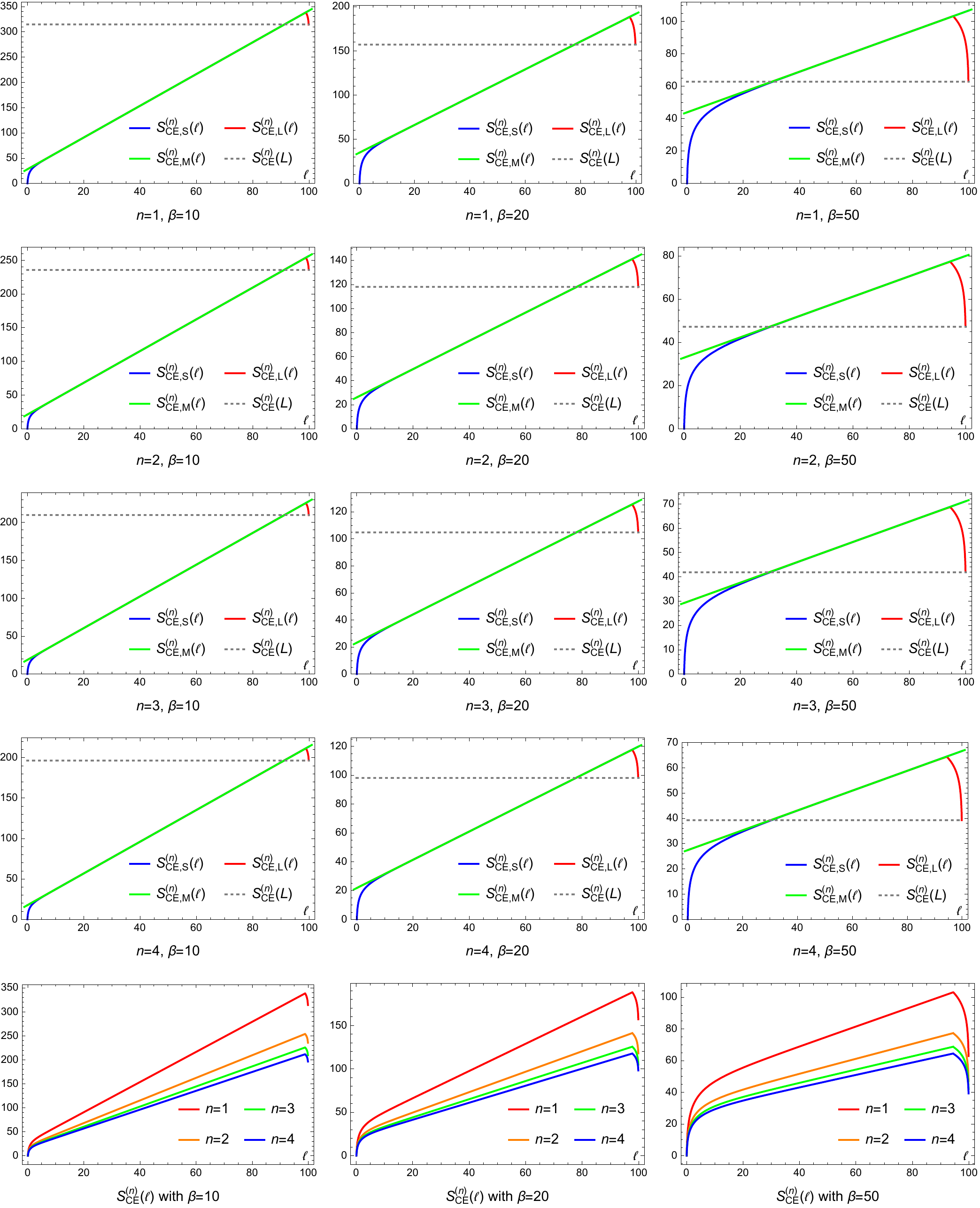}
  \caption{In the first four rows, we plot the R\'enyi entropy of short interval (\ref{SceS}), medium interval (\ref{SceM}), long interval (\ref{SceL}) for canonical ensemble state, and we also plot the R\'enyi entropy of the total system (\ref{Scetot}). In the last row, we plot the piecewise R\'enyi entropy (\ref{SceSML}). To draw the figures, we have set the central charge $c=30$, the length of the entire system $L=100$, and UV cutoff of the CFT $\e=0.1$.}\label{Sce}
\end{figure}

In summary, the piecewise R\'enyi entropy for the canonical ensemble state is
\be \label{SceSML}
S_{\CE}^{(n)}(\ell) = \lt\{
\ba{ll}
S_{\CE,\rS}^{(n)}(\ell) & \e <\ell<\ell^\CE_1 \\
S_{\CE,\rM}^{(n)}(\ell) & \ell^\CE_1<\ell<\ell^\CE_2 \\
S_{\CE,\rL}^{(n)}(\ell) & \ell^\CE_2<\ell<L-\e
\ea
\rt.\!\!\!\!.
\ee
Note that in the above we have put back the UV cutoff $\e$.
We plot the short interval, medium interval, long interval, and piecewise R\'enyi entropies of the canonical ensemble state (\ref{SceS}), (\ref{SceM}), (\ref{SceL}), (\ref{SceSML}) in Figure~\ref{Sce}.

\section{Microcanonical ensemble state} \label{secME}

We then consider a large energy density microcanonical ensemble state with energy $E=\f{\pi c L}{6\l^2}$. For the short interval we use the OPE of twist operators \cite{Headrick:2010zt,Calabrese:2010he,Rajabpour:2011pt,Chen:2013kpa} to calculate R\'enyi entropy, and the result can be written as a sum of the products of one-point functions \cite{Chen:2016lbu,Lin:2016dxa,He:2017vyf,He:2017txy}.
It was recently shown in \cite{Guo:2018djz} that in the thermodynamic limit and with the identification $\b=\l$, the canonical and microcanonical ensemble states have the same one-point functions so that the resultant short interval R\'enyi entropies are the same as long as the short interval expansion converges.
We thus get the short interval R\'enyi entropy in microcanonical ensemble state
\be \label{SmeS}
S_{\ME,\rS}^{(n)}(\ell)= \f{c(n+1)}{6n} \log \Big( \f{\l}{\pi\e} \sinh \f{\pi\ell}{\l} \Big).
\ee
Note that it is only valid for $\e < \ell \lesssim \l$.

For a long interval, we can still use the OPE of twist operators \cite{Chen:2014ehg,Chen:2015kua}, and we give the calculation details  in Appendix~\ref{appdetails}. The long interval R\'enyi entropy in microcanonical ensemble state is
\be \label{SmeL}
S_{\ME,\rL}^{(n)}(\ell)= \f{c(n+1)}{6n} \log \Big( \f{\l}{\pi\e} \sinh \f{\pi(L-\ell)}{\l} \Big)
                        +\f{\pi c L}{3\l}.
\ee
It is only valid for $L - \l \lesssim \ell < L-\e$.
As $\ell$ approaches $L-\e$,  it approaches the R\'enyi entropy of the total system
\be \label{Smetot}
S_{\ME}^{(n)}(L) = \f{\pi cL}{3\l}.
\ee
Unlike the canonical ensemble case (\ref{Scetot}), the RHS of (\ref{Smetot})  does not depend on the R\'enyi index $n$.

The R\'enyi entropy of a medium interval can be calculated holographically as in \cite{Dong:2018lsk}. The backreacted geometry $\mB_n(\b_n,A)$ caused by the cosmic brane for a microcanonical ensemble state of dual CFT is approximately the same as $\mB_n(\b,A)$ for a  canonical ensemble state  with the parameter $\b_n$ given by \cite{Dong:2018lsk}
\be
\b_n = \l \sr{1-\f{\ell}{L}+\f{\ell}{n^2L}}.
\ee
Using this fact and the result for a canonical ensemble state (\ref{tdSceM}), we can get the medium interval refined R\'enyi entropy for the corresponding microcanonical ensemble state
\be\label{MErefine}
\td S_{\ME,\rM}^{(n)}(\ell) = \f{c}{3n} \log \f{\l \sr{1-\f{\ell}{L}+\f{\ell}{n^2L}}}{2\pi\e} + \f{\pi c \ell}{3n\l \sr{1-\f{\ell}{L}+\f{\ell}{n^2L}}}.
\ee
The medium interval R\'enyi entropy in microcanonical ensemble state can thus be obtained from (\ref{MErefine}),
\bea \label{SmeM}
&& S_{\ME,\rM}^{(n)}(\ell) = \f{c(n+1)}{6n} \log \f{\l}{2\pi\e}
                        - \f{c n L}{12(n-1)\ell} \Big( 1-\f{\ell}{L}+\f{\ell}{n^2L} \Big) \log \Big( 1-\f{\ell}{L}+\f{\ell}{n^2L} \Big)
                        - \f{c(n+1)}{12n} \nn\\
&& \phantom{S_{\ME,\rM}^{(n)}(\ell) =}
                        + \f{\pi c nL}{3(n-1)\l} \Big( 1-  \sr{1-\f{\ell}{L}+\f{\ell}{n^2L}} \Big).
\eea
Again, comparing with the results in \cite{Dong:2018lsk}, i.e. $\td S_{\ME,\rM}^{(n)}(\ell)$ and $S_{\ME,\rM}^{(n)}(\ell)$ in (\ref{ScenMSmeMnD}), we find some additional terms.
This is similar to the canonical ensemble case, as we discussed below (\ref{tdSceM}).
In the refined R\'enyi entropy (\ref{MErefine}), the extra term is due to the edge effect of the cosmic brane, and this also leads to extra terms in the R\'enyi entropy (\ref{SmeM}). Our results are consistent with and generalize the results in \cite{Dong:2018lsk}. The validity of the results in \cite{Dong:2018lsk} requires
\be
\f{L}{\l} \gg \log \f{\l}{\e}.
\ee
Our results with the extra terms do not need such a requirement, and the validity region of the results is enlarged.
As we will see in section~\ref{secdist}, the extra terms we find are subleading and do not affect the distinguishabilities of the canonical and microcanonical states from the R\'enyi entropy.

\begin{figure}[htpb]
  \centering
  \includegraphics[width=0.9\textwidth]{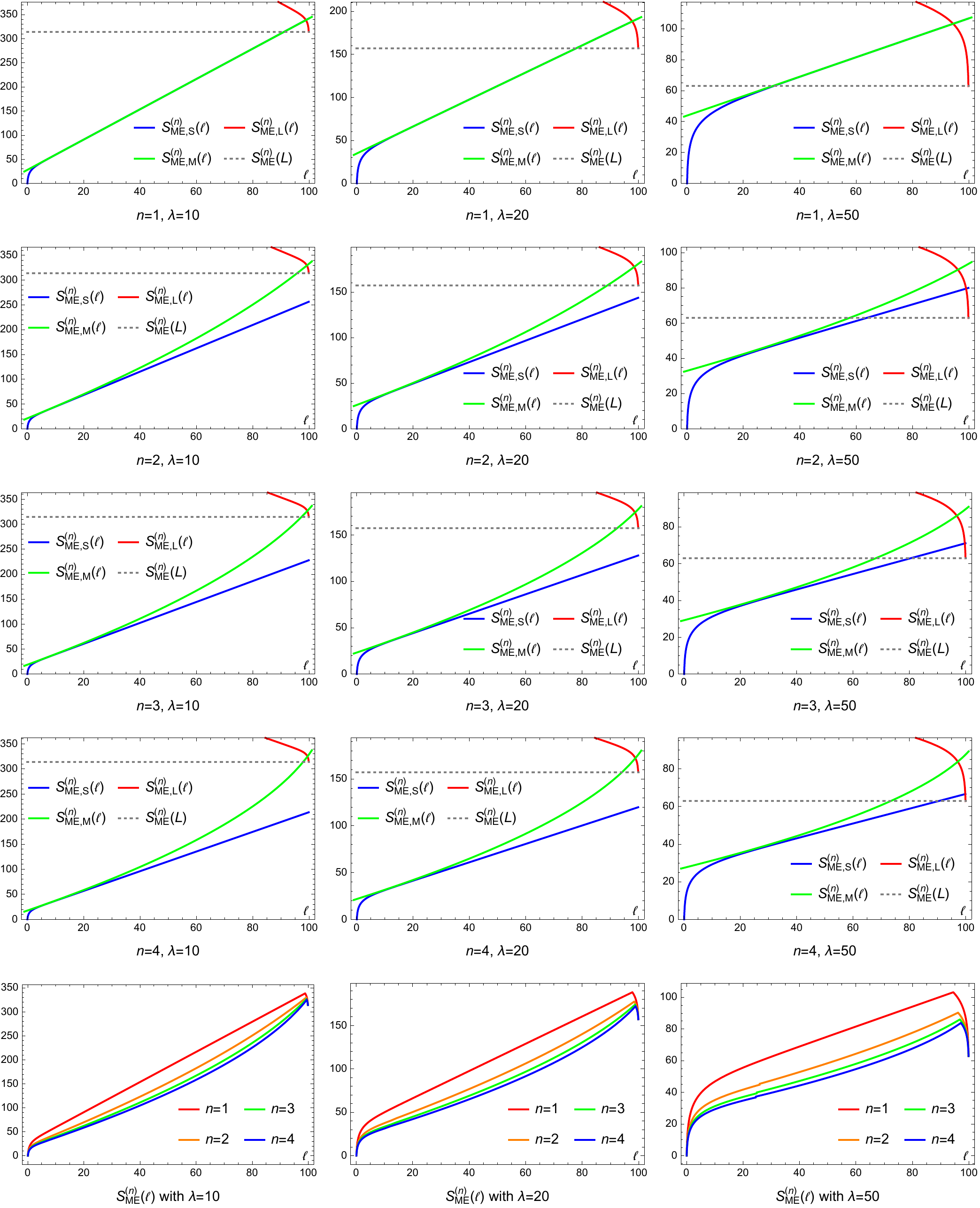}
  \caption{In the first four rows, we plot the R\'enyi entropy of short interval (\ref{SmeS}), medium interval (\ref{SmeM}), long interval (\ref{SmeL}) for microcanonical ensemble state in the thermodynamic limit, and we also plot the R\'enyi entropy of the total system (\ref{Smetot}). In the last row, we plot the piecewise R\'enyi entropy (\ref{SmeSML}). To draw the figures, we have set $c=30$, $L=100$, and $\e=0.1$.}\label{Sme}
\end{figure}

 Similar to the canonical ensemble case, due to the limited regimes of validity for the short and long interval formulas (\ref{SmeS}) and (\ref{SmeL}), the medium interval formula (\ref{SmeM}) should also have a validity region, say $\ell^\ME_1 \lesssim \ell \lesssim \ell^\ME_2$ for some critical lengths $\ell_{1,2}^\ME$. Thus, the piecewise R\'enyi entropy for a microcanonical ensemble state is
\be \label{SmeSML}
S_{\ME}^{(n)}(\ell) = \lt\{
\ba{ll}
S_{\ME,\rS}^{(n)}(\ell) & \e<\ell<\ell^\ME_1 \\
S_{\ME,\rM}^{(n)}(\ell) & \ell^\ME_1<\ell<\ell^\ME_2 \\
S_{\ME,\rL}^{(n)}(\ell) & \ell^\ME_2<\ell<L-\e
\ea
\rt.\!\!\!\!.
\ee
However, we do not know how to obtain the precise forms of the critical lengths $\ell^\ME_{1,2}$.
Given $n$ and $\l$, we can get the approximate value of $\ell^\ME_1$ by requiring $S_{\ME,\rS}^{(n)}(\ell^\ME_1) \approx S_{\ME,\rM}^{(n)}(\ell^\ME_1)$, and the approximate $\ell^\ME_2$ by $S_{\ME,\rM}^{(n)}(\ell^\ME_2) \approx S_{\ME,\rL}^{(n)}(\ell^\ME_2)$.
As expected, both $\ell^\ME_1$ and $L-\ell^\ME_2$ are the same order of $\l$.
In fact, by setting $\l=\b$, $\ell^\ME_{1,2}$ are, respectively, close to $\ell^\CE_{1,2}$.
%The piecewise R\'enyi entropy for a typical microcanonical ensemble state is shown in Figure~\ref{SceSmeSpeX1}.
We plot the short interval, medium interval, long interval, and piecewise R\'enyi entropies in the microcanonical ensemble state (\ref{SmeS}), (\ref{SmeM}), (\ref{SmeL}), (\ref{SmeSML}) in the Figure~\ref{Sme}.

\section{Primary excited state} \label{secPE}

We finally consider a large energy density primary excited state with conformal weights $h=\bar h =\f{c}{24} ( \f{L^2}{\m^2} + 1 )$ and energy $E=\f{\pi c L}{6\m^2}$.
The short interval R\'enyi entropy for a primary excited state was calculated in \cite{Lashkari:2016vgj,Lin:2016dxa,He:2017vyf,He:2017txy}, and in the thermodynamic limit it is
\bea \label{SpeS}
&& S_{\PE,\rS}^{(n)}(\ell) = \frac{c (n+1)}{6 n} \log\f\ell\e
                            +\frac{\pi ^2 c (n+1) \ell ^2}{36 n \mu ^2}
                            -\frac{\pi ^4 c (n+1) (n^2+11) \ell ^4}{12960 n^3 \mu^4} \nn\\
&& \phantom{S_{\PE,\rS}^{(n)}(\ell) =}
                            -\frac{\pi ^6 c (n+1) (n^2-4) (n^2+47) \ell^6}
                                  {2449440 n^5 \mu ^6}
                            +\frac{\pi ^8 c (n+1) \ell^8}
                                  {1175731200 (5 c+22) n^7 \mu ^8}
                             [ c (13 n^6-1647 n^4\nn\\
&& \phantom{S_{\PE,\rS}^{(n)}(\ell) =}
                               +33927n^2-58213)-88 (n^2-4) (n^2-9) (n^2+119) ]
                            +O\Big(\f\ell\m\Big)^{10}.
\eea
No closed form of the short interval R\'enyi entropy in primary excited state is known, and we use the  above expansion as an approximation. The expansion breaks down  as $\ell$ approaches $\m$.

It was argued in \cite{Lu:2017tbo,Dong:2018lsk} that the medium interval R\'enyi entropy in the primary excited state is the same as that in the microcanonical ensemble state as long as $\ell < L/2$. If so, this would lead to the medium interval R\'enyi entropy in primary excited state
\bea \label{SpeM}
&& S_{\PE,\rM}^{(n)}(\ell) \overset{?}{=} \f{c(n+1)}{6n} \log \f{\m}{2\pi\e}
                        - \f{c n L}{12(n-1)\ell} \Big( 1-\f{\ell}{L}+\f{\ell}{n^2L} \Big) \log \Big( 1-\f{\ell}{L}+\f{\ell}{n^2L} \Big)
                        - \f{c(n+1)}{12n} \nn\\
&& \phantom{S_{\PE,\rM}^{(n)}(\ell) \overset{?}{=}}
                        + \f{\pi c nL}{3(n-1)\m} \Big( 1-  \sr{1-\f{\ell}{L}+\f{\ell}{n^2L}} \Big).
\eea
We use the symbol ``$\overset{?}{=}$'' to remind the reader that there is no rigorous justification.
We do not know whether the conjecture is true at the leading order of $c$, but in the next section we will show that even if it is true at the leading order of $c$ there must be some corrections at order $O(c^0)$.
In \cite{Dong:2018lsk} the author commented that the conjecture may likely fail for 2D CFTs, which are special compared to their higher dimensional cousins because of the infinite number of commuting conserved quantum Korteweg-de Vries charges \cite{Sasaki:1987mm,Bazhanov:1994ft}. In fact, a different result of the R\'enyi entropy in the primary excited state from (\ref{SpeM}) was obtained in \cite{Faulkner:2017hll}, and one can also see an earlier proposal in \cite{Dymarsky:2016ntg}. The medium interval R\'enyi entropy in primary excited state is an open question, as emphasized in both \cite{Faulkner:2017hll} and \cite{Dong:2018lsk}, however we will plot the figures using the conjecture (\ref{SpeM}).

\begin{figure}[htpb]
  \centering
  \includegraphics[width=0.9\textwidth]{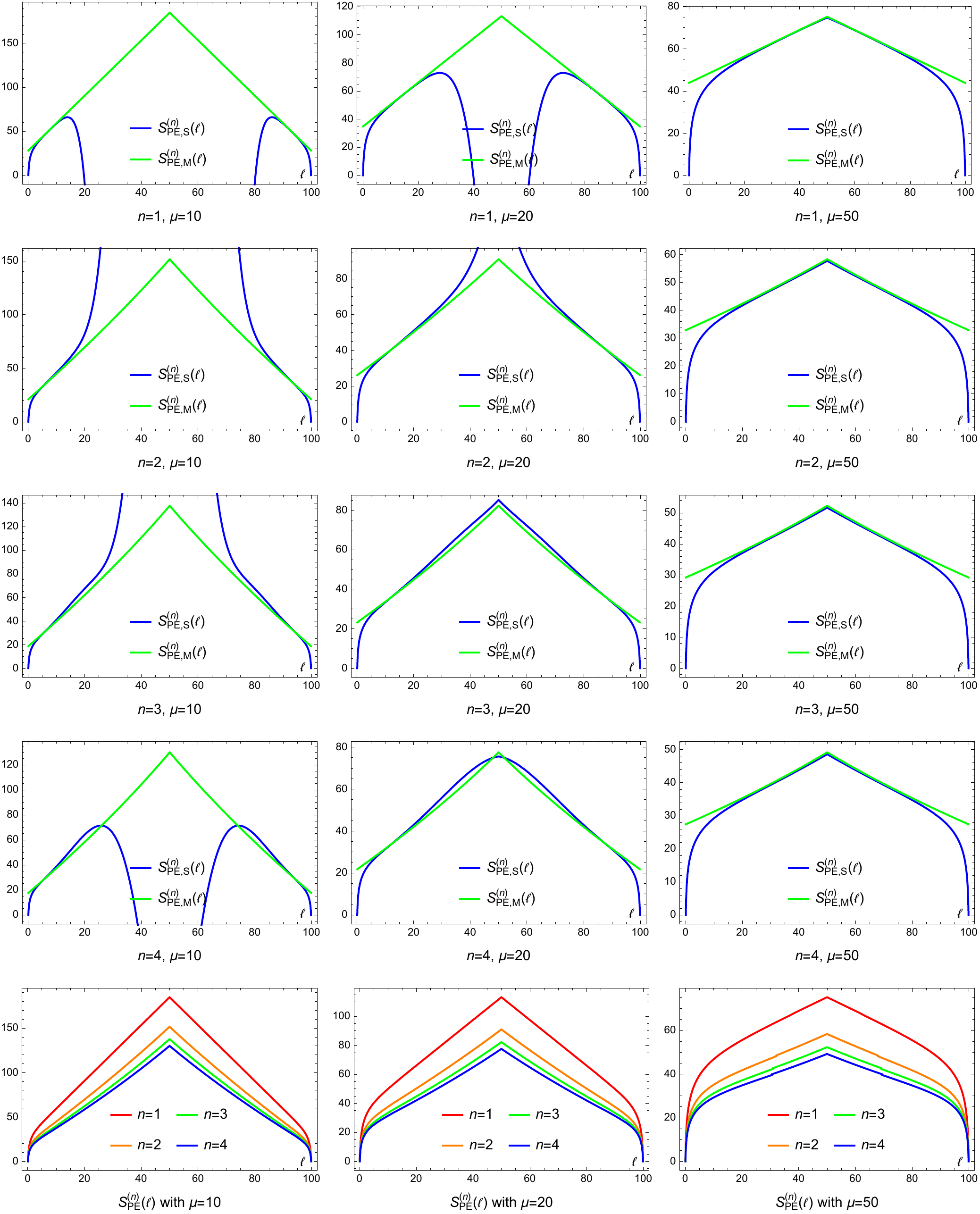}
  \caption{In the first four rows, we plot the R\'enyi entropy of short interval (\ref{SpeS}) and medium interval (\ref{SpeM}), and for the long interval $L/2 < \ell < L$ we use $S_{\PE}^{(n)}(\ell)=S_{\PE}^{(n)}(L-\ell)$. In the last row, we plot the piecewise R\'enyi entropy (\ref{SpeSML}).
  We remind the reader that the medium interval R\'enyi entropy for the primary excited state (\ref{SpeM}) is a conjecture.
  To draw the figures, we have set $c=30$, $L=100$, and $\e=0.1$.}\label{Spe}
\end{figure}

There should exist a critical length $\ell_\PE$ so that the short interval formula (\ref{SpeS}) holds only for $\e<\ell\lesssim\ell_\PE$  and the medium interval formula   (\ref{SpeM}) holds only for $\ell_\PE\lesssim\ell<L/2$. Given $n$, $\m$, we can determine the approximate value of $\ell_\PE$ from $S_{\PE,\rS}^{(n)}(\ell_\PE) \approx S_{\PE,\rM}^{(n)}(\ell_\PE)$.  By setting $\l=\b=\mu$, we find $\ell_\PE$ is close to $\ell_1^\CE$ and $\ell_1^\ME$. For the long interval regime $L/2<\ell<L-\e$, we can use $S_{\PE}^{(n)}(\ell)=S_{\PE}^{(n)}(L-\ell)$ to get the long interval R\'enyi entropy. In summary, we get the piecewise R\'enyi entropy for a primary excited state
\be \label{SpeSML}
S_{\PE}^{(n)}(\ell) = \lt\{
\ba{ll}
S_{\PE,\rS}^{(n)}(\ell) & \e<\ell<\ell_\PE \\
S_{\PE,\rM}^{(n)}(\ell) & \ell_\PE<\ell<L/2 \\
S_{\PE}^{(n)}(L-\ell) & L/2<\ell<L-\e
\ea
\rt.\!\!\!\!.
\ee
%The piecewise R\'enyi entropy for a typical primary excited state is shown in Figure~\ref{SceSmeSpeX1}.
We plot the short interval, medium interval, and piecewise R\'enyi entropies in the primary excited state (\ref{SpeS}), (\ref{SpeM}), (\ref{SpeSML}) in the Figure~\ref{Spe}.

\section{Distinguishabilities of various states} \label{secdist}

\begin{figure}[htpb]
  \centering
  \includegraphics[width=0.9\textwidth]{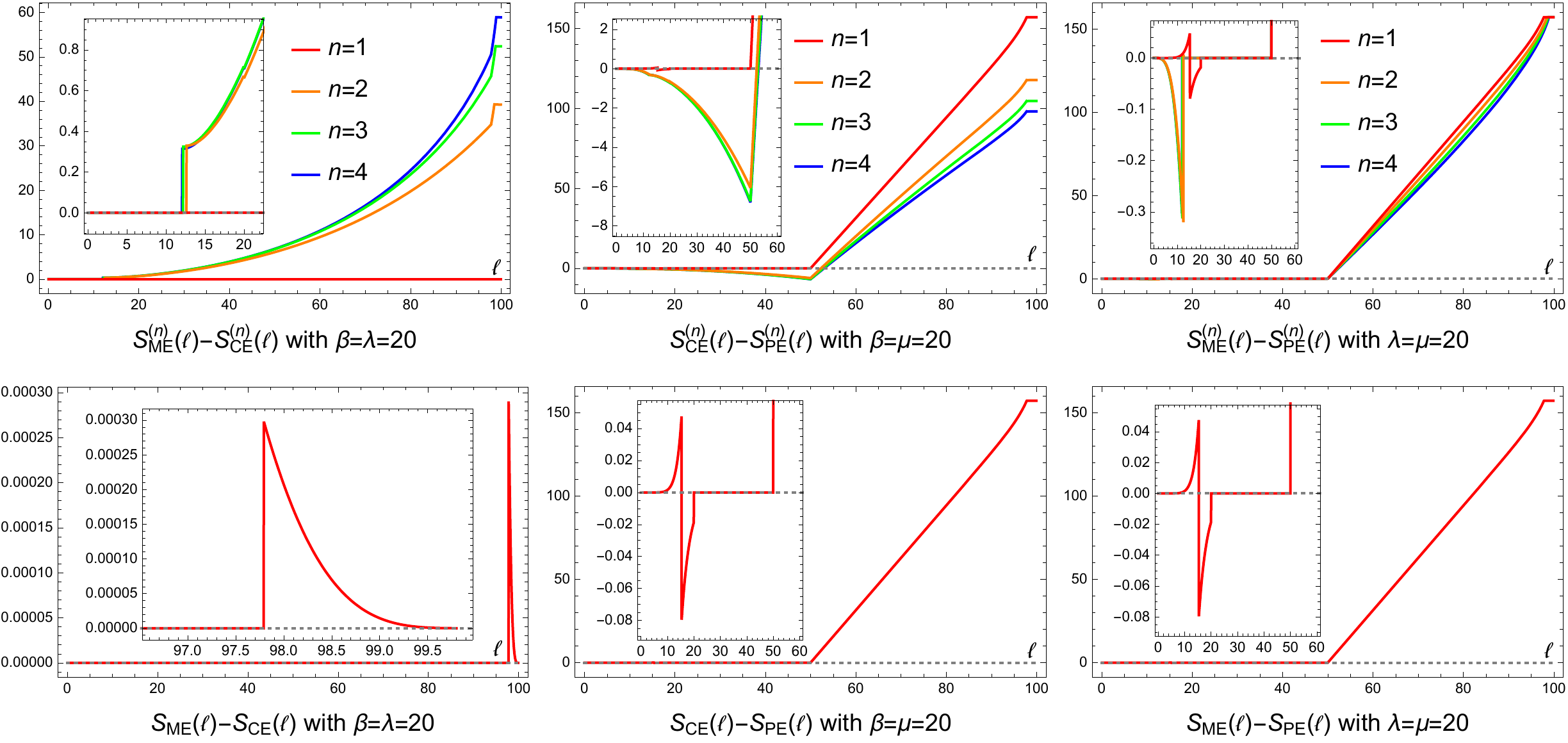}
  \caption{The differences of R\'enyi entropies (in upper row) and entanglement entropies (in lower row) in the canonical ensemble, microcanonical ensemble, and primary excited states from (\ref{SceSML}), (\ref{SmeSML}), and (\ref{SpeSML}). As we cannot calculate the precise various critical lengthes, i.e. $\ell_{1,2}^\CE$, $\ell_{1,2}^\ME$, $\ell_\PE$, the figures around these critical lengthes are just suggestive.
  We also remind the reader that the medium interval R\'enyi entropy in the primary excited state (\ref{SpeM}) is a conjecture, and so the corresponding R\'enyi entropy and entanglement entropy differences are also conjectures and thus suggestive.
  Especially, the medium interval regime of entanglement entropy difference $S_{\CE}(\ell) - S_{\PE}(\ell)$ (middle of the 2nd row) should be nonvanishing.
  To draw the figures we have set $c=30$, $L=100$.}\label{SceSmeSpeX}
\end{figure}

As stated in \cite{Dong:2018lsk}, to investigate ETH, it is interesting to use the R\'enyi entropy and entanglement entropy to distinguish various large energy density states.
We use the results (\ref{SceSML}), (\ref{SmeSML}), and (\ref{SpeSML}) to calculate the differences of the the R\'enyi entropies and entanglement entropies among the canonical ensemble, microcanonical ensemble, and primary excited states, and plot the results in Figure~\ref{SceSmeSpeX}.
Note that since we cannot calculate the precise various critical lengthes, i.e. $\ell_{1,2}^\CE$, $\ell_{1,2}^\ME$, $\ell_\PE$, the figures around these critical lengthes are just suggestive.
We also remind the reader that the medium interval R\'enyi entropy in the primary excited state (\ref{SpeM}) is a conjecture, and so the corresponding R\'enyi entropy and entanglement entropy differences are also conjectures and thus suggestive.

%\begin{figure}[htpb]
%  \centering
%  \includegraphics[width=0.9\textwidth]{SceSmeSpeX1.pdf}
%  \caption{The piecewise R\'enyi entropies of an arbitrary length interval in, respectively, the canonical ensemble state (\ref{SceSML}) (left), microcanonical ensemble state (\ref{SmeSML}) (middle), and primary excited state (\ref{SpeSML}) (right).
%  To draw the figures we have set the central charge $c=30$, the entire system length $L=100$, and the UV cutoff $\e=0.1$.}\label{SceSmeSpeX1}
%\end{figure}

   %Even assuming the conjecture (\ref{SpeM}) holds and thus $S^{(n)}_{\ME,\rM}(\ell) - S^{(n)}_{\PE,\rM}(\ell)=0$ at the leading order of $c$, in the following we will argue that this will not hold for the sub-leading order of $c$. First,
    The leading order $c$ part of the entanglement entropy for the primary excited state with length $0 < \ell < L/2$  was calculated in \cite{Asplund:2014coa,Caputa:2014eta}, and it is the same as the entanglement entropy in canonical ensemble state with the identification $\b=\m$
\be
S_\PE(\ell) = \f{c}{3} \log \Big( \f{\m}{\pi\e} \sinh \f{\pi\ell}{\m} \Big) + O(c^0).
\ee
No closed form of the $1/c$ corrections is known, and it was calculated by short interval expansion to $\ell^8$ in \cite{He:2017vyf,He:2017txy} and to $\ell^{12}$ in \cite{Guo:2018pvi}.
Let  us denote  the reduced density matrices of the interval $A$ for the canonical ensemble and primary excited states by $\r_\CE(\ell)$ and $\r_\PE(\ell)$, respectively. Using the fact that in the thermodynamic limit the modular Hamiltonian of $\r_\CE(\ell)$ is a local integral of the energy density \cite{Wong:2013gua,Cardy:2016fqc}, the relative entropy will be reduced to the difference of the entanglement entropies, i.e., \cite{Lashkari:2016vgj}
\be\label{ReEE}
S(\r_\PE(\ell) \| \r_\CE(\ell)) = \tr_A[ \r_\PE(\ell) \log \r_\PE(\ell) ] - \tr_A[ \r_\PE(\ell) \log \r_\CE(\ell) ]
                                = S_\CE(\ell) - S_\PE(\ell).
\ee
Note that this formula holds as long as  $\ell$ is not comparable to $L$. Moreover, if we set $\l=\m$ and use the result in (\ref{SceS}) and (\ref{SpeS}), we can get the entanglement entropy difference of the short interval
\be
S_{\CE,\rS}(\ell) - S_{\PE,\rS}(\ell)=S(\r_\PE(\ell) \| \r_\CE(\ell))
                                = \frac{121 \pi ^8 c \ell ^8}{510300 (5 c+22) \l ^8} + O\Big(\f{\ell}{\l}\Big)^{10}.
\ee
Since the entanglement entropy difference $S_\CE(\ell) - S_\PE(\ell)$ inherits the non-negativity and monotonicity of the corresponding relative entropy \cite{nielsen2010quantum}, this yields that the medium interval entanglement entropy difference must be nonvanishing and be of at least order $O(c^0)$ in the large $c$ limit. Thus the entanglement entropy of a medium interval can distinguish the canonical ensemble and primary excited states {at least at the order of $O(c^0)$.

   Finally, combining the above statement with the fact that the medium interval entanglement entropies of the canonical and microcanonical ensemble states are the same at all orders of $c$, we can conclude that the entanglement entropy of a medium interval can also distinguish the microcanonical ensemble and primary excited states at least at the order of $O(c^0)$. Since the entanglement entropy is just a special case of R\'enyi entropy, this conclusion should also hold for R\'enyi entropy, i.e., $S^{(n)}_{\ME,\rM}(\ell) - S^{(n)}_{\PE,\rM}(\ell)=O(c^0)$.
   %if assuming the conjecture (\ref{SpeM}) holds at the leading order of $c$.

We summarize the distinguishabilities of the canonical ensemble, microcanonical ensemble, and primary excited states in Table~\ref{tab}.
Especially,  the R\'enyi entropy cannot distinguish the canonical and microcanonical ensemble states for a short interval, but can distinguish the two states for a medium interval and a long interval at the leading order of $c$.
At the leading order of $c$ the entanglement entropy cannot distinguish the canonical and microcanonical ensemble states for any length interval, but the difference of entanglement entropy between the two states would appear with $1/c$ corrections for a long interval.
Both R\'enyi entropy and entanglement entropy can easily distinguish the thermal states and the primary excited state for a long interval.
The R\'enyi entropy is more powerful than the entanglement entropy to distinguish different states.
Our findings are consistent with and generalize the holographic medium interval results in \cite{Dong:2018lsk},  and are also consistent with the conjecture in \cite{Dymarsky:2016ntg}.

\begin{table}[htpb]
  \centering
\begin{tabular}{|c|c|c|c|c|c|c|c|c|}\hline
%\begin{tabular}{|c|c|p{0.7cm}<{\centering}|p{0.7cm}<{\centering}|p{0.7cm}<{\centering}|p{0.7cm}<{\centering}|}\hline
  \mlr{2}{states} & \mlr{2}{interval} & \mlc{2}{c|}{leading order of $c$}                                                                                     & \mlc{2}{c|}{all orders of $c$} \\ \cline{3-6}
                  &                   & entanglement                                        & R\'enyi                                         & entanglement                             & R\'enyi \\ \hline
  canonical       & short             & \texttimes\cite{Guo:2018djz}                                & \texttimes\cite{Guo:2018djz}                            & \texttimes\cite{Guo:2018djz}                     & \texttimes\cite{Guo:2018djz} \\ \cline{2-6}
  VS              & medium            & \texttimes\cite{Dong:2018lsk}                               & \checked\cite{Dong:2018lsk}                             & \texttimes\cite{Dong:2018lsk}                    & \checked\cite{Dong:2018lsk} \\ \cline{2-6}
  microcanonical  & long              & \texttimes                                                  & \checked                                                & \checked                                         & \checked \\ \hline
  canonical       & short             & \texttimes\cite{Asplund:2014coa,Caputa:2014eta}             & \checked\cite{Lashkari:2016vgj,Lin:2016dxa}             & \checked\cite{He:2017vyf,He:2017txy}             & \checked\cite{Lashkari:2016vgj,Lin:2016dxa} \\ \cline{2-6}
  VS              & medium            & \texttimes(?)\cite{Lu:2017tbo,Dong:2018lsk}                 & \checked(?)\cite{Lu:2017tbo,Dong:2018lsk}               & \checked                                         & \checked \\ \cline{2-6}
  primary         & long              & \checked                                                    & \checked                                                & \checked                                         & \checked \\ \hline
  microcanonical  & short             & \texttimes\cite{Asplund:2014coa,Caputa:2014eta,Guo:2018djz} & \checked\cite{Lashkari:2016vgj,Lin:2016dxa,Guo:2018djz} & \checked\cite{He:2017vyf,He:2017txy,Guo:2018djz} & \checked\cite{Lashkari:2016vgj,Lin:2016dxa,Guo:2018djz} \\ \cline{2-6}
  VS              & medium            & \texttimes(?)\cite{Lu:2017tbo,Dong:2018lsk}                 & \texttimes(?)\cite{Lu:2017tbo,Dong:2018lsk}             & \checked                                         & \checked \\ \cline{2-6}
  primary         & long              & \checked                                                    & \checked                                                & \checked                                         & \checked \\ \hline
\end{tabular}
  \caption{The distinguishabilities of the canonical ensemble, microcanonical ensemble, and primary excited states for a short, medium, and long interval in terms of the entanglement entropy and R\'enyi entropy.
  We mark ``\checked"  for distinguishable states, and mark ``\texttimes" otherwise.
  For some cases we give the references where the results were firstly derived or could be easily inferred from.
  Note that for ``canonical VS primary'' and ``microcanonical VS primary'', we refer to $\m \lesssim \ell <L/2$ for a medium interval and refer to $L/2 < \ell <L$ for a long interval.
  The medium interval R\'enyi entropy in the primary excited state (\ref{SpeM}) is a conjecture, and we mark ``(?)" for the corresponding cases.
  The other cases are derived in this paper.}\label{tab}
\end{table}

In Table~\ref{tab}, there is a \textit{puzzle}, and it is related to that the medium interval R\'enyi entropy for the primary excited state (\ref{SpeM}) is a conjecture in \cite{Lu:2017tbo,Dong:2018lsk}.
Generally, the states become more distinguishable as the length of the interval increases.
When two states can be distinguished for a short interval, one may expect that they can also be distinguished for a medium interval.
In Table~\ref{tab} there is one case that this rule does not apply, which is using the leading order R\'enyi entropy to distinguish the microcanonical ensemble and primary excited states.
Generally, there is no theorem to guarantee that the R\'enyi entropy difference must be nondecreasing with respect to $\ell$.
As we have discussed above, the conjecture (\ref{SpeM})  at least at the order of $O(c^0)$, but we cannot conclude whether it is true at the leading order of $c$.

In the thermodynamics limit, using directly the definition of the relative entropy and the density matrices of the entire system, with $\b=\l$ we get the vanishing relative entropy of the entire system in the microcanonical and canonical ensemble states, i.e.,
\be\label{EEentire}
S(\r_\ME(L)\|\r_\CE(L)) = 0.
\ee
%which can be derived by using the definition of relative entropy
%\be
%S(\r_\ME(L)\|\r_\CE(L))=tr[\r_\ME(L) \log \r_\ME(L)]\\ -tr[\r_\ME(L) \log \r_\CE(L)]\ee
% and $\r_\CE(L)=e^{-\beta H}/tr[e^{-\beta H}]$.
From (\ref{Scetot}) and (\ref{Smetot}), we see that two infinitely closed states in terms of the relative entropy can have different R\'enyi entropies. From the monotonicity of the relative entropy \cite{nielsen2010quantum}, we further get for an arbitrary length interval
\be
S(\r_\ME(\ell)\|\r_\CE(\ell)) = 0.
\ee
The reduced density matrices of the medium and long intervals in the microcanonical and canonical ensemble states are the extra examples that two infinitely closed states in terms of the relative entropy have different R\'enyi entropies.
From the modular Hamltonian \cite{Lashkari:2016vgj,Wong:2013gua,Cardy:2016fqc}, we get for the short and medium intervals
\be
S_\CE(\ell) = S_\ME(\ell).
\ee
This is a direct proof in the CFT that the canonical and microcanonical ensemble states have the same short and medium interval entanglement entropies.
This is consistent with the holographic results in \cite{Dong:2018lsk} and the previous results in this paper.

With $\b=\m$, it is also easy to get the relative entropy of the entire system in the primary excited and canonical ensemble states
\be
S(\r_\PE(L)\|\r_\CE(L)) = \f{\pi c L}{3\b}.
\ee
This is consistent with the result in this paper. We cannot extract any additional useful information from this relative entropy.

\section{Conclusion and discussion} \label{secdisc}

Motivated by the recent work \cite{Dong:2018lsk}, we investigate the distinguishabilities of three high energy density states in a 2D large $c$ CFT using the R\'enyi entropy and entanglement entropy. Our work provides an up-to-date picture of distinguishing the different types of thermal states or primary states at various length scales of the subsystem. To achieve this, we make some new calculations and collect some known results in literature for comparison with our findings. Our new findings are the R\'enyi entropy of the long interval  and the R\'enyi entropy of the medium interval in the microcanonical ensemble state, which complements and enlarges the validity region of the holographic result in \cite{Dong:2018lsk}. We proved generally in CFT that the canonical and microcanonical ensemble states have the same short and medium interval entanglement entropies.
While the long interval entanglement entropies in the canonical and microcanonical ensemble states are the same at the leading order of $c$, we showed explicitly that they are different at order $O(c^0)$.
With some argument we conclude the conjecture for the medium interval R\'enyi entropy in the primary excited state should be corrected at least at order $O(c^0)$.

If ETH works, it should apply not only to the primary states but also the descendant states in the 2D large $c$ CFT. The short interval entanglement entropy for some special descendant states was recently studied in \cite{Guo:2018fnv,Guo:2018pvi}, and they generally behave differently from the canonical ensemble, microcanonical ensemble, or primary excited states, especially when the descendant states are highly excited above the corresponding primary states. It would be nice if R\'enyi entropy and entanglement entropy for general descendant states could be calculated and compared with the thermal states.

The nonvanishing differences of the short interval R\'enyi entropies and entanglement entropies for the primary excited state and the canonical ensemble or microcanonical ensemble state indicate the failure of ETH in context of canonical ensemble or microcanonical ensemble.
A possible solution is to consider the generalized Gibbs ensemble (GGE) \cite{Rigol:2006}, instead of the canonical ensemble or microcanonical ensemble. The ETH in context GGE has been studies in \cite{He:2017vyf,Basu:2017kzo,He:2017txy,Lashkari:2017hwq,Dymarsky:2018lhf,Maloney:2018hdg,Maloney:2018yrz,Dymarsky:2018iwx,Brehm:2019fyy,Dymarsky:2019etq}, and whether it works is still an open question.

 In investigations of the R\'enyi entropies, we have considered three regimes according to the length of the interval. This is necessary as in different regimes the R\'enyi entropies are derived in different methods. In (1+1)D CFTs to calculate the R\'enyi entropy of one interval one needs to evaluate the two-point function of local twist operators in  CFT$^n$, which are the heavy operators with the conformal dimension $h_n\sim c$.  Of course it is also interesting to use the two-point function of more general local operators to study the distinguishabilities of the three kinds of high energy density states\footnote{We thank the anonymous referee for pointing this out to us. The two-point function of the stress tensor was used recently in \cite{Datta:2019jeo} to find typical states that is accurately thermal.}.
For a short interval, we can use the OPE to reduce the two-point function into the combination of one-point functions, then one can see that the two-point function cannot distinguish the canonical and microcanonical ensemble states as the two states have exactly the same one-point functions in the thermodynamic limit. It would be nice to see if a two-point function with a larger length can distinguish the two states.

\section*{Acknowledgments}

We would like to thank Anatoly Dymarsky for helpful discussions.
We thank the anonymous referees of JHEP for helpful suggestions.
WZG is supported by the National Center of Theoretical Science (NCTS).
FLL is supported by Taiwan Ministry of Science and Technology through Grant No.~103-2112-M-003-001-MY3.
JZ acknowledges support from ERC under Consolidator grant number 771536 (NEMO).

\appendix

\section{R\'enyi mutual information}\label{appMI}

On a complex plane the R\'enyi mutual information of two disjoint intervals $A=[z_1,z_2]$ and $B=[z_3,z_4]$ is defined as
\be
I_{A,B}^{(n)} = S_A^{(n)} + S_B^{(n)} - S_{A \cup B}^{(n)}.
\ee
It is a function of the cross ratio $x=\f{(z_1-z_2)(z_3-z_4)}{(z_1-z_3)(z_2-z_4)}$, and one can write it as $I_n(x)$.
For the 2D large $c$ CFT, we only include the contributions from the vacuum conformal family.
The R\'enyi mutual information satisfies the property \cite{Headrick:2010zt}
\be
I_n(x) = \f{c(n+1)}{6n}\log\f{x}{1-x} + I_n(1-x).
\ee
No closed form of $I_n(x)$ is known, and for a small $x \ll 1$ it has been calculated up to order $x^{8}$ \cite{Hartman:2013mia,Faulkner:2013yia,Barrella:2013wja,Chen:2013kpa,Chen:2013dxa}.
One can see the result in \cite{Chen:2013dxa}.
We plot it in Figure~\ref{MI}.

\begin{figure}[htpb]
 \centering
  \includegraphics[width=0.4\textwidth]{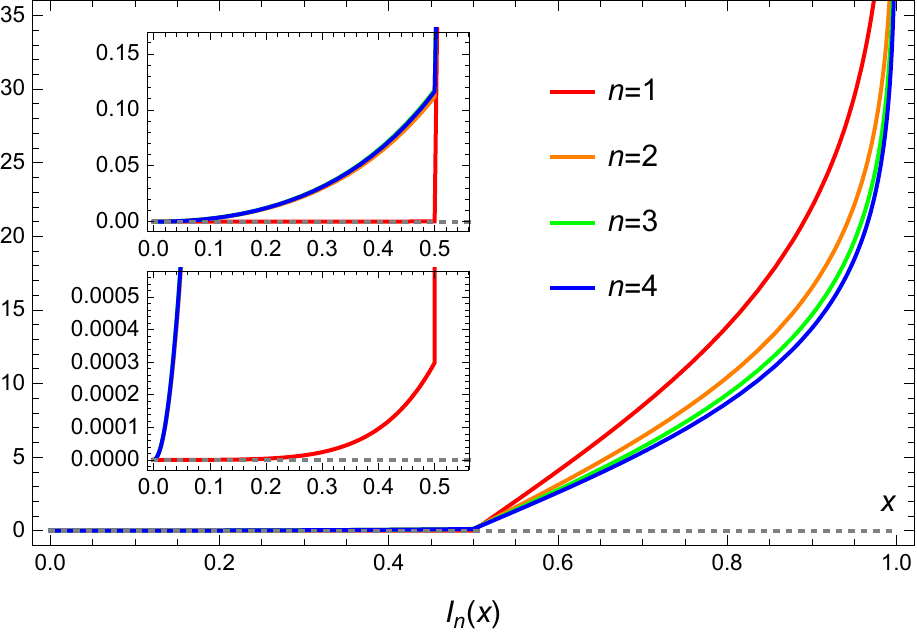}
  \caption{The R\'enyi mutual information of two disjoint intervals on a complex plane with cross ratio $x$ in the 2D large $c$ CFT. To draw the figure we have set $c=30$.}\label{MI}
\end{figure}

\section{Derivation of long interval R\'enyi entropy in microcanonical ensemble state}\label{appdetails}

We derive the long interval R\'enyi entropy (\ref{SmeL}) for the high energy density microcanonical ensemble state with energy $E=\f{\pi c L}{6\l^2}$ in the thermodynamic limit in the 2D large $c$ CFT.

The density matrix of the entire system of the microcanonical ensemble state and the partition function are
\be
\r(E) = \f{1}{\O(E)} \sum_i \d(E-E_i) |i \rag \lag i |, ~~
\O(E) = \sum_i \d(E-E_i).
\ee
The density matrix of canonical ensemble state with inverse temperature $\b$ and the canonical ensemble partition function are
\be
\r(\b) = \f{1}{Z(\b)} \sum_i \ep^{-\b E_i} |i \rag \lag i |, ~~
Z(\b) = \sum_i \ep^{-\b E_i}.
\ee
There is the relation
\be
\O(E) = \int_{\g-\ii\inf}^{\g+\ii\inf}\f{d\b}{2\pi\ii} \ep^{\b E} Z(\b).
\ee
We have a short interval $A$ with length $\ell$ and its complement $\bar A$, and a long interval with length $L-\ell$.
Note that $S_{\ME,\rS}^{(n)}(\ell)=S_A^{(n)}$ and
$S_{\ME,\rL}^{(n)}(L-\ell)=S_{\bar A}^{(n)}$.

From OPE of twist operators \cite{Headrick:2010zt,Calabrese:2010he,Rajabpour:2011pt,Chen:2013kpa}, for the short interval reduced density matrix $\r_A(E)=\tr_{\bar A}\r(E)$ we get \cite{Chen:2016lbu,Lin:2016dxa,He:2017vyf,He:2017txy}
\be \label{trArAEn}
\tr_A [\r_A(E)^n] = c_n \Big( \f{\ell}{\e} \Big)^{-4h_\s}
                     \Big(
                       1 +
                       \sum_{k=1}^n \sum_{\{\mX_1, \cdots, \mX_k\}}\ell^{\D_{\mX_1}+\cdots\D_{\mX_k}}
                                                                   b_{\mX_1\cdots\mX_k}
                                                                   \lag \mX_1 \rag_{\r(E)}\cdots\lag \mX_k \rag_{\r(E)}
                     \Big),
\ee
with $h_\s=\f{c(n^2-1)}{24n}$  and the one-point function of a general quasiprimary operator $\mX$ in the 2D CFT being defined as
\be
\lag \mX \rag_{\r(E)} = \f{1}{\O(E)}\sum_i\d(E-E_i)\lag i|\mX|i\rag.
\ee
The coefficients $b_{\mX_1\cdots\mX_k}$ were defined in \cite{Chen:2016lbu} and are related to the OPE coefficients of the twist operators, and their explicit forms are not important to us in this paper.
In the thermodynamic limit the one-point function w.r.t. microcanonical ensemble state is the same as the expectation value w.r.t. the canonical ensemble state with inverse temperature $\b$ that equals $\l$ \cite{Guo:2018djz}
\be
\lag \mX \rag_{\r(E)} =\lag \mX \rag_{\r(\b)}|_{\b \to \l}.
\ee
The short interval expansion converges for $0<\ell \lesssim \b=\l$.
We derive the short interval R\'enyi entropy for microcanonical ensemble state (\ref{SmeS}).

For the long interval reduced density matrix $\r_{\bar A}(E)=\tr_{A}\r(E)$ we get \cite{Chen:2014ehg,Chen:2015kua}
\bea \label{trAbrAbEn}
&& \tr_{\bar A} [\r_{\bar A}(E)^n] = c_n \Big( \f{\ell}{\e} \Big)^{-4h_\s} \Big( \f{\d(0)}{\O(E)} \Big)^{n-1}
                     \Big[
                       1 + \sum_{k=1}^n \sum_{\{\mX_1, \cdots, \mX_k\}} \Big( \ell^{\D_{\mX_1}+\cdots\D_{\mX_k}} b_{\mX_1\cdots\mX_k} \\
&& ~~~~~~\times
                       \f{1}{\d(0)^{k-1}\O(E)}
                       \sum_{i_1, \cdots, i_k}
                       \d(E-E_{i_1}) \cdots \d(E-E_{i_k})
                       \lag i_1 | \mX_1 | i_2 \rag
                       \cdots
                       \lag i_{k-1} | \mX_{k-1} | i_k \rag
                       \lag i_k | \mX_k | i_1 \rag \Big)
                     \Big]. \nn
\eea
We define
\bea
&& \cJ_{\mX_1\cdots\mX_k}(E;\o_1, \cdots, \o_{k-1}) = \f{1}{\d(0)^{k-1}\O(E)}
                                                   \sum_{i_1, \cdots, i_k}
                                                   \big[
                                                   \d(E-E_{i_1}) \\
&& ~~~~~~\times
                                                   \d(\o_1-E_{i_1}+E_{i_2})\cdots \d(\o_{k-1}-E_{i_{k-1}}+E_{i_k})
                                                   \lag i_1 | \mX_1 | i_2 \rag
                                                   \cdots
                                                   \lag i_{k-1} | \mX_{k-1} | i_k \rag
                                                   \lag i_k | \mX_k | i_1 \rag \big], \nn
\eea
and it can be expressed as a series of Fourier transformations plus an inverse Laplace transformation of a multi-point function in the canonical ensemble state
\bea
&& \cJ_{\mX_1\cdots\mX_k}(E;\o_1, \cdots, \o_{k-1}) = \f{1}{\d(0)^{k-1}\O(E)}
   \int_{\g-\ii\inf}^{\g+\ii\inf}\f{d\b}{2\pi\ii}
   \int_{-\inf}^{+\inf}\f{dt_1}{2\pi}
   \cdots
   \int_{-\inf}^{+\inf}\f{dt_{k-1}}{2\pi} \\
&& \phantom{\cJ_{\mX_1\cdots\mX_k}(E;\o_1, \cdots, \o_{k-1}) =} \times
   \big[
   \ep^{\b E -\ii ( \o_1t_1 + \cdots + \o_{k-1}t_{k-1})}
   Z(\b) \lag \mX_1(t_1) \cdots \mX_{k-1}(t_{k-1})\mX_k(0) \rag_{\r(\b)} \big]. \nn
\eea
Note that $t_1,\cdots,t_{k-1}$ are the Minkowski time, and in the multi-point function w.r.t. the canonical ensemble state the quasiprimary operators $\mX_1,\cdots,\mX_k$ are inserted at the same spatial position $x=0$ but different temporal positions.
For $\o_1 = \cdots = \o_{k-1} =0$, it is just the second line of (\ref{trAbrAbEn})
\bea \label{JX1dddXkE0ddd0}
&& \cJ_{\mX_1\cdots\mX_k}(E;0, \cdots, 0) = \f{1}{\d(0)^{k-1}\O(E)}
   \int_{\g-\ii\inf}^{\g+\ii\inf}\f{d\b}{2\pi\ii}
   \int_{-\inf}^{+\inf}\f{dt_1}{2\pi}
   \cdots
   \int_{-\inf}^{+\inf}\f{dt_{k-1}}{2\pi} \nn\\
&& \phantom{\cJ_{\mX_1\cdots\mX_k}(E;0, \cdots, 0) =} \times
   \big[
   \ep^{\b E}
   Z(\b) \lag \mX_1(t_1) \cdots \mX_{k-1}(t_{k-1})\mX_k(0) \rag_\b \big].
\eea

We have used the Dirac delta function for the energy $\d(E-E_i)$, and we can also define the Dirac delta function for the scaling dimension $\td\d(\D-\D_i)$. From $E=\f{2\pi}{L}( \D -\f{c}{12} )$, we get
\be
\d(E-E_i) = \f{L}{2\pi}\td\d(\D-\D_i).
\ee
Especially, we have
\be
\d(0) = \f{L}{2\pi}\td\d(0).
\ee
Note that $L \to \inf$ in the thermodynamic limit and $\td\d(0)=\inf$. We find that the only term in (\ref{JX1dddXkE0ddd0}) that survive in the thermodynamic limit is
\bea
&& \cJ_{\mX_1\cdots\mX_k}(E;0, \cdots, 0) = \f{1}{\d(0)^{k-1}\O(E)}
   \int_{\g-\ii\inf}^{\g+\ii\inf}\f{d\b}{2\pi\ii}
   \int_{-\inf}^{+\inf}\f{dt_1}{2\pi}
   \cdots
   \int_{-\inf}^{+\inf}\f{dt_{k-1}}{2\pi}
   \big[
   \ep^{\b E} \nn\\
&& \phantom{\cJ_{\mX_1\cdots\mX_k}(E;\o_1, \cdots, \o_{k-1}) =} \times
   Z(\b)
   \lag \mX_1(t_1) \rag_{\r(\b)}
   \cdots
   \lag \mX_{k-1}(t_{k-1}) \rag_{\r(\b)}
   \lag \mX_k(0) \rag_{\r(\b)} \big].
\eea
Noting that the one-point functions w.r.t. the canonical ensemble state are constants, we get
\be
\cJ_{\mX_1\cdots\mX_k}(E;0, \cdots, 0) = \lag \mX_1 \rag_{\r(\b)} \cdots \lag \mX_k \rag_{\r(\b)}|_{\b \to \l}.
\ee
We obtain the relation of the partition functions
\be
\tr_{\bar A} [\r_{\bar A}(E)^n] = \Big( \f{\d(0)}{\O(E)} \Big)^{n-1} \tr_A [\r_A(E)^n],
\ee
and the relation of the R\'enyi entropies
\be
S_{\bar A}^{(n)} = S_A^{(n)} + \f{\pi c L}{3\l}.
\ee
We have discarded the divergent term $-\log\d(0)$.
We derive the long interval R\'enyi entropy for microcanonical ensemble state (\ref{SmeL}).

\providecommand{\href}[2]{#2}\begingroup\raggedright\endgroup

%\bibliographystyle{D:/00.zbib/utphys}
%\bibliographystyle{D:/00.zbib/jhep-z}
%\bibliography{D:/00.zbib/2019,D:/00.zbib/2018,D:/00.zbib/1900,D:/00.zbib/1960,D:/00.zbib/1970,D:/00.zbib/1980,D:/00.zbib/1990,D:/00.zbib/1995,D:/00.zbib/1996,D:/00.zbib/1997,D:/00.zbib/1998,D:/00.zbib/1999,D:/00.zbib/2000,D:/00.zbib/2001,D:/00.zbib/2002,D:/00.zbib/2003,D:/00.zbib/2004,D:/00.zbib/2005,D:/00.zbib/2006,D:/00.zbib/2007,D:/00.zbib/2008,D:/00.zbib/2009,D:/00.zbib/2010,D:/00.zbib/2011,D:/00.zbib/2012,D:/00.zbib/2013,D:/00.zbib/2014,D:/00.zbib/2015,D:/00.zbib/2016,D:/00.zbib/2017,D:/00.zbib/book,D:/00.zbib/work,D:/00.zbib/thesis}

\end{document}